\documentstyle[twocolumn,aps,psfig]{revtex}
\newcommand{\ignore}[1]{}
\begin{document}
\def\be{\begin{equation}}
\def\ee{\end{equation}}
\def\ba{\begin{eqnarray}}
\def\ea{\end{eqnarray}}

\title{Spin structure of many-body systems with two-body random interactions}
\author{Lev Kaplan$^1$, Thomas Papenbrock$^1$, and Calvin W. Johnson$^2$}
\address{$^1$Institute for Nuclear Theory,
University of Washington, Seattle, Washington 98195}
\address{$^2$Department of Physics and Astronomy, Louisiana State University, 
Baton Rouge, Louisiana 70803}
\maketitle
\begin{abstract}
We investigate the spin structure of many-fermion systems with a 
spin-conserving two-body random interaction. We find a strong dominance
of spin-0
ground states and considerable correlations between energies and wave functions
of low-lying states with different spin,
but no indication of pairing. The spectral
densities exhibit spin-dependent shapes and widths, and depend on the
relative strengths
of the spin-0 and spin-1 couplings in the two-body random matrix. The spin
structure of low-lying states can largely be explained analytically.
\end{abstract}
\pacs{PACS numbers: 05.30.-d, 21.60.Cs, 73.23.-b, 24.60.Lz}

\section{Introduction}

Random matrix theory has been widely used to model the behavior of complex
quantum systems~\cite{Brody,GMW}. Of particular importance are two-body random
matrix ensembles (TBRE), which were introduced into nuclear physics three
decades ago~\cite{FW70,BF71} and recently have received a considerable amount of
renewed interest in the study of mesoscopic
systems~\cite{Shepel,Silvestrov,Carlos,Izrailev,Alhassid,Jacquod} and
nuclei~\cite{Zelevinsky,JBD,Sahu,BFP,Kota,BF,JBDT,KP,BF2}.
This is partly due to the
observation of regular structures in the
low-lying states of various nuclear models
with random two-body interactions. As examples we mention the observed ground
state dominance of $J^P=0^+$ states, the occurrence of pairing (energy) gaps in
random shell models~\cite{JBD}, and the existence of vibrational and rotational
bands in random interacting boson models~\cite{BFP}.  While these observations
concern the low-lying levels of many-body systems, similar regularities
also have been
reported in the wave function structure. These include signatures of
phonon collectivity relating states of different angular momentum~\cite{JBD},
strong pair-transfer amplitudes between ground states of nuclei with different
mass numbers~\cite{JBDT}, strong electromagnetic transitions in bands of random
interacting boson models~\cite{BF,BF2}, deviations from random matrix theory in
transition strength distributions for embedded Gaussian ensembles~\cite{Sahu},
and regularities in wave functions of spin systems with random
interactions~\cite{KP}. These observations indicate that some typical
properties of interacting many-body systems are determined simply by the
presence of a rotationally invariant two-body force that allows for transitions
between single-particle Fock space states.

In this article we aim at an understanding of some of the reported
regularities. We are particularly interested in the quantum numbers of the
ground state and in correlations between energies and wave functions of
low-lying states.  For this purpose we consider a system of $N$ spin-${1\over
2}$ fermions on $M$ orbitals, subject to a spin-conserving random two-body
interaction. This system is simpler than the nuclear models studied so far, and
allows us to derive analytical results. It may be seen as a model for
mesoscopic systems like quantum dots.
 
This article is divided as follows. In the following Section we focus on the
width and shape of spectral densities in sectors of definite spin for two
models of spin systems with random interactions, and derive analytical results.
Special emphasis is placed on understanding the spin structure of the spectrum
close to the ground state.  Our analysis extends and complements previous work
in nuclei~\cite{BFP} and mesoscopic systems~\cite{Jacquod}. In
Section~\ref{seccorrelpairing}
we investigate correlations between low-lying states, and consider
pairing properties.  Finally we give a summary.

\section{Spin Structure}
\label{secspinstruct}

\subsection{The model}

We begin with a very simple model of random spin-conserving
two-body interactions in a many-body
system~\cite{JBD,Sahu,BFP,Kota,BF,JBDT}:
\begin{equation}
H=C_0\sum_{\alpha,\alpha' \; {\rm spin} 0} H_{\alpha \alpha'}
A^\dagger_\alpha A_{\alpha'} + C_1 \sum_{\beta,\beta' \; {\rm spin} 1}
\tilde{H}_{\beta \beta'} A^\dagger_\beta A_{\beta'}\,,
\label{twobodyham}
\end{equation}
where $\alpha$ and $\alpha'$ each denotes a pair
of two spin$-{1 \over 2}$ fermions in a $J=0$ state, while $\beta$ and $\beta'$
each denotes a pair of fermions in a $J=1$ state. The two-body
states can of course be labeled by the individual
orbitals occupied by the two particles.
Thus, for a system with $M$ single-particle orbitals, we have $M(M-1)/2$
pairs with total spin $J=1$,
enumerated as $A_{\beta_{ij}}=(a_{i \downarrow}a_{j \uparrow} +
a_{i \uparrow}a_{j \downarrow})/{\sqrt 2}$, where $a_j$ annihilates a single particle state, and the single-particle
indices are $1\le i < j\le M$. [Here we are enumerating two-body
$J=1$ states with $J_z=0$ only; obviously there are equal numbers
of $J_z=\pm 1$ two-body states which by rotational symmetry must couple
in precisely the same way.] Similarly, there are $M(M-1)/2$
pairs with total spin $J=0$, having the form
$A_{\alpha_{ij}}=(a_{i \downarrow}a_{j \uparrow} -
a_{i \uparrow}a_{j \downarrow})/{\sqrt 2}$, and an additional $M$ pairs
where a single site is doubly occupied: $A_{\alpha_{ii}}=a_{i \downarrow}
a_{i \uparrow}$. We will assume throughout that the two kinds of $J=0$
states interact equally strongly among themselves and with each other; this
assumption does not substantially affect any of the results to be presented
in this paper. The effect of diagonal spin-spin interactions in a spin system
with random wave functions has been investigated in ref.~\cite{Jacquod}.  

Within each two-body spin sector, then, the couplings
$H_{\alpha \alpha'}$ and $\tilde{H}_{\beta \beta'}$
are assumed to be governed by a GOE random-matrix ensemble,
\begin{eqnarray}
\left \langle H_{\alpha \alpha'}^2 \right \rangle &=&
 1 +\delta_{\alpha \alpha'} \nonumber
\\
\left \langle \tilde{H}_{\beta \beta'}^2 \right 
\rangle &=& 1 +\delta_{\beta \beta'}  \,,
\end{eqnarray}
where $\langle \ldots \rangle$ indicates an ensemble average.  We note that the
two matrices $H_{\alpha \alpha'}$ and $\tilde H_{\beta \beta'}$
are each taken to be
real and symmetric, consistent with time reversal invariance; the model can
easily be generalized to the case where an external magnetic field is present,
so that the matrices must be taken from a GUE rather than a GOE ensemble,
or from
an intermediate ensemble~\cite{Brody,GMW,BFP}.  Finally, we notice that in the
model (Eq.~\ref{twobodyham}), couplings in different two-body spin sectors are
taken to be completely independent of one another; in
Section~\ref{secvijkl}
we will discuss an alternative random-coupling model where this is not
the case. In Section~\ref{seccorrelpairing}
we will see what effects on spin spectra
may be obtained by adding a one-body term to the two-body random
Hamiltonian. Note that the presence of a strong
one-body term enhances the likelyhood
of finding a spin-0 ground state, due to spin-degeneracy of the one-body
spectrum.

\subsection{Preliminary results}

We may now consider numerically the spectra of $N-$fermion systems interacting
via the random two-body interaction of the form (Eq.~\ref{twobodyham}). Due
to an approximate particle-hole symmetry of this system, maximum density is
obtained near half-filling, $N=M$, and we may restrict ourselves to the
parameter range $2 \le N \le M$. The preponderance of $S=0$ many-body ground
states, which has been observed previously in shell model calculations,
is seen to be an extremely strong effect in this simple random interaction
model. For $N=4$ fermions on $M=4$ sites, with an exclusively spin-$0$
interaction ($C_0=1$, $C_1=0$), we find for an ensemble of $2000$ systems
that the ground state has total spin $S=0$ in $99.8\%$ of the cases. For
an exclusively spin-$1$ interaction ($C_1=1$, $C_0=0$), the ground state is
observed to have total spin $S=0$ for $99.75\%$ of the systems in the
ensemble (a result which is
statistically indistinguishable from the previous case).
This is despite the
fact that in this system $S=0$ many-body states comprise only about $56\%$
of the total number of states. For $N=4$ particles on $M=5$ sites, all $2000$
computed ground states were observed to have total spin $S=0$, independent of
whether a pure
$J=0$ or $J=1$ two-body coupling was used. Very similar results
are obtained at least until a system size of $M=12$, beyond which point
the size of the many-body basis makes the diagonalization of a large ensemble
of Hamiltonians prohibitively expensive. For $N=6$ particles on $M=6,$
$7,$ and $8$ sites, we once again observe an almost $100\%$ prevalence of
$S=0$ ground states.

How may we understand these initially surprising findings within the context of
a completely random interaction model? We begin by considering the shape of the
many-body density of states $\rho_S(E)$ in sectors of total many-body
spin $S$. It is
known that for $N \gg 2$ particles with a two-body interaction, the moments of
the spectral density move away from those associated with a semicircle law, and
approach those corresponding to a Gaussian shape~\cite{Gervots,Mon}. Then, the
ensemble-averaged spectra $\langle \rho_S(E) \rangle$, being centered at $E=0$
because of the symmetry of the Hamiltonian, must be described simply by their
squared widths, ${\rm Tr}_S H^2$ (where ${\rm Tr}_S$ denotes throughout
the sum over all
basis states in the sector of total spin $S$, divided by the number of basis
states in that sector).  Our naive expectation (assuming that the spectrum can
be described by a Gaussian shape well into its tail) is that ground state
behavior should be associated with the relative widths of the different spin
spectra, a preponderance of $S=0$ ground states arising from a large value of
${\rm Tr}_0 H^2$ as compared to widths in the other sectors~\cite{Jacquod,BFP}.
Indeed, fixing the number of particles at $N=4$, our numerical results show
that for $4 \le M \le 10$ orbitals the relation ${\rm Tr}_0 H^2 > {\rm Tr}_1
H^2 > {\rm Tr}_2 H^2$ is almost always observed (independent of the relative
values of $C_0$ and $C_1$), consistent with $S=0$ ground state dominance.  For
$M=11$ and $12$, however, when taking an exclusively $J=1$ coupling, this
pattern is reversed, with the widths of the higher-spin spectra beginning to
exceed those of the $S=0$ spectrum. This shows that the width analysis alone is
not sufficient to explain numerical results for the ground state
behavior. However, these findings do
suggest than an interesting transition may be taking place as we approach the
dilute regime, which we shall now proceed to understand analytically.

\subsection{Width analysis}
\label{secwidthgen}

The many-body
basis states of the system may be classified by the total spin $S$ and the
number $D$ of orbitals that are doubly occupied, $0 \le D \le N/2$. We easily
see that in the dilute limit ($M \to \infty$ for fixed $N$), $D=0$ states
will dominate: a typical $S=0$ or $S=1$ basis state in this limit is given
for even $N$ by
\begin{eqnarray}
|\Psi\rangle &=& 
{1 \over 2^{N/2}}\left (a^\dagger_{i\downarrow} a^\dagger_{j \uparrow} \pm
a^\dagger_{i\uparrow} a^\dagger_{j \downarrow} \right)
\nonumber \\ & \times & \prod_{z=1}^{N/2-1}
\left (
a^\dagger_{k_z\downarrow} a^\dagger_{l_z \uparrow} -
a^\dagger_{k_z\uparrow} a^\dagger_{l_z \downarrow}\right )|0\rangle \,,
\label{typecmany}
\end{eqnarray}
where the plus (minus) sign produces an $S=1$ ($S=0$) $N-$body state,
the single-particle orbitals $i$, $j$, $k_1$, $l_1$, \ldots, $k_{N/2-1}$,
$l_{N/2-1}$ are all taken to be distinct, and $|0\rangle$ denotes the vacuum.
However, at finite densities we must consider 
basis states in subsectors of different $D$ for any given value of $S$, as
these might in general have spectra of different shape and width. It is a
straightforward, though somewhat tedious,
counting problem to compute the ensemble-averaged squared
energy width
$\langle \Psi |H^2| \Psi\rangle$, for a typical basis
state $|\Psi \rangle$ of given quantum numbers $S$ and $D$,
when acted on
by the Hamiltonian (Eq.~\ref{twobodyham}). For each two-particle pair
$\alpha'$ or $\beta'$ of the appropriate spin that may be annihilated, one
simply has to count contributions from all pairs $\alpha$
or $\beta$ that may be created in its
place, while keeping track of terms  that must be added up coherently.
So, for example, for a pure $J=0$ coupling ($C_0=1$, $C_1=0$), one obtains
\begin{eqnarray}
\label{j0gen}
{\rm Tr}_0 H^2 &=& { N^2(2M-N)^2\over 64}+ {2M^2N +MN^2 -N^3\over 16} \nonumber
\\ &+&{6MN+N^2 +2N^2 \left[ \left( {D \over N/2} \right)^2 -
\left( {D \over N/2} \right )\right ]\over 16} + {N \over 2} \nonumber \\
{\rm Tr}_1 H^2 &=& {\rm Tr}_0 H^2 - {2M^2-2MN+N^2 \over 4} \nonumber \\
&-& {3M-N \over 2}-1
\,.
\end{eqnarray}

In Eq.~\ref{j0gen}, terms are grouped and ordered by decreasing
power in the number of orbitals $M$
assuming finite density ($N \sim M$). In particular, we notice
that the leading $O(M^4)$ term is invariant under particle-hole symmetry,
$N \to 2M-N$. This symmetry is broken at $O(1/M)$ because of new terms
that arise in the  Hamiltonian (Eq.~\ref{twobodyham}) when one commutes
annihilation operators $A$ and creation operators $A^\dagger$ corresponding to
pairs having at least one orbital in common. We notice also that the effect
of double degeneracy ($D>0$) is initially to increase the width slightly, and
then restore it to the original value when $D$ reaches its maximal value of
$N/2$. However, the effect of double degeneracy appears only at $O(1/M^2)$,
which we will see below to be small compared with the level spacing near the
ground state, in the many-body limit. Thus, even for a non-dilute system
($N \sim M$), we are justified in studying the spectra of $D=0$ states and
taking their widths as proxies for the entire spectrum of that spin, even
for the
purpose of determining behavior near the edge of the spectrum (where the level
spacings are largest).
In Section~\ref{secpairing} we will consider the possible relevance of $D>0$
contributions to wave function structure near the ground state. 
We find there that basis states with high (or low) double occupancy
have almost equal overlaps with
eigenstates of all energies, and have no special tendency to contribute
to the low-lying part of the spectrum. Thus it appears that,
to leading order, double occupancy
effects are not important to understanding the spin structure of the spectrum
for a pure two-body interaction.

\subsection{The four-body system}
\label{secwidth}

The width expression of Eq.~\ref{j0gen} is rather unwieldy, as it consists
of $14$ monomial terms, and so we
begin our detailed
analysis with the special case of $N=4$ particles; as we shall see below,
all the interesting qualitative behavior of the many-body system is already
present in this simple case. 
In the $S=0$
sector we have $2{M \choose 4}$ independent basis states of type $D=0$,
having the form
\begin{equation}
{1 \over 2}\left(a^\dagger_{i\downarrow} a^\dagger_{j \uparrow} -
a^\dagger_{i\uparrow} a^\dagger_{j \downarrow} \right)\left(
a^\dagger_{k\downarrow} a^\dagger_{l \uparrow} -
a^\dagger_{k\uparrow} a^\dagger_{l \downarrow}\right)|0\rangle \,,
\label{type0c}
\end{equation}
(where $i$, $j$, $k$, and $l$ all represent different orbitals),
$M{M-1 \choose 2}$ basis states of type $D=1$, and ${M \choose 2}$
basis states of type $D=2$. Similarly, in the $S=1$ sector we have
$3{M \choose 4}$ independent basis states of type $D=0$, of the form
\begin{equation}
{1 \over 2}\left (a^\dagger_{i\downarrow} a^\dagger_{j \uparrow} +
a^\dagger_{i\uparrow} a^\dagger_{j \downarrow}\right )\left (
a^\dagger_{k\downarrow} a^\dagger_{l \uparrow} -
a^\dagger_{k\uparrow} a^\dagger_{l \downarrow}\right )|0\rangle \,,
\label{type1c}
\end{equation}
and $M{M-1 \choose 2}$ basis states of type $D=1$. Finally, in the $S=2$
sector we only have the ${M \choose 4}$ basis states of type $D=0$. 
As the system becomes dilute ($M \to \infty$), type
$D=0$ states always dominate, and the total numbers of states in
the spin $0$, $1$, and $2$ sectors approach the ratio $2:3:1$. [As we
mentioned above, even for finite $M$, the inclusion of $D>0$ states does not
significantly affect the behavior of the spin spectra, leading to corrections
that are small compared with the level spacing.]

What then causes the preponderance of $S=0$ ground states in this system?
Using $D=0$ basis states, one obtains the following expressions for the widths
(of course the $C_0^2$ part of these results is merely a special case of
Eq.~\ref{j0gen}):
\begin{eqnarray}
{\rm Tr}_0 H^2 &=& C_0^2\left ({3 \over 2}M^2 -{3 \over 2}M+3\right)+
C_1^2\left ( {3 \over 2}M^2+{3 \over 2}M+3\right) \nonumber \\
{\rm Tr}_1 H^2 &=& C_0^2 \left (M^2 -M\right)+
C_1^2\left ( 2M^2-4M+4 \right)\,.
\label{traces4}
\end{eqnarray}
The leading $O(M^2)$ terms in the above expressions are easiest to understand.
We simply need to take the number of spin-$J$ pairs in a four-body state of
total spin $S$ (Eq.~\ref{type0c} or Eq.~\ref{type1c}), and multiply by the
number of orbital pairs to which the two particles may jump ($\approx
M^2/2$). So, for example, in a state of total spin $S=0$ (Eq.~\ref{type0c}),
the $ij$ and $kl$ pairs are in pure $J=0$ combinations, while the remaining $4$
pairs are uncorrelated, i.e. $1/4$ of them have $J=0$ and $3/4$ have
$J=1$. This leads to a total of $3$ pairs each of $J=0$ and $J=1$, explaining
the two ${3 \over 2} M^2$ terms in the first line of Eq.~\ref{traces4}. In a
state of total spin $S=1$ (Eq.~\ref{type1c}), the only difference is that the
$ij$ pair is now in a $J=1$ combination, so there are only $2$ pairs with
$J=0$ and $4$ pairs with $J=1$, leading to the $C_0^2M^2$ and $2C_1^2M^2$ terms
in the second line of Eq.~\ref{traces4}. We see already from this very simple
argument that, due to a relatively larger number of $J=0$ pairs in a spin-$0$
state and a relatively larger number of $J=1$ pairs in a spin-$1$ state, in the
$M \to \infty$ limit we always have ${\rm Tr}_0 H^2 > {\rm Tr}_1 H^2$ if $C_0 >
C_1$ and ${\rm Tr}_0 H^2 < {\rm Tr}_1 H^2$ if $C_0 < C_1$. The ratio of the
widths in the dilute limit approaches a constant,
\begin{equation}
{{\rm Tr}_0 H^2 \over {\rm Tr}_1 H^2} \approx {3C_0^2+3C_1^2 \over
2C_0^2 +4C_1^2}\,,
\end{equation}
while the ratio of the many-body level spacing near the
ground state, $\rho_S^{-1}(E_{gs})$, to the width and thus to the width
difference is
becoming small. Because
the spectral shapes $\rho_{0,1}(E)$ approach
a Gaussian form in the $M\to \infty$ limit, we expect this width difference
to dominate the ground-state behavior, and to see $100\%$ $S=0$ ground
states for a strong $C_0$ coupling, and $0\%$ for a stronger $C_1$ coupling.

\subsection{The $C_0=C_1$ special case}

The $M \to \infty$ behavior for $C_0=C_1$ is more subtle and requires an
analysis of the subleading $O(M)$ terms in Eq.~\ref{traces4}. To compute
these we need to count more carefully the number of places $\alpha$ or $\beta$
to which the
two interacting particles may jump, given that two
of the orbitals remain occupied
by the two `bystanders' during the interaction.
In the case of a $J=1$ pair, we also
need to treat correctly the interference between $J_z=0,\pm 1$ terms. The end
result is that overall, after adding together $J=0$ and $J=1$ couplings, the
$S=1$ width is suppressed at $O(M)$ compared with the $S=0$ width, as a simple
physical argument shows. In Eq.~\ref{type0c},
the $jk$ pair (or any one of the other
$3$ analogous pairs) may interact either in a $J=0$ or $J=1$ combination, with
the final state leaving $j$ doubled up with $i$ and with
$k$ ending up on any of the
remaining $M-1$ sites. For an initial state of the
form Eq.~\ref{type1c}, this process is
not allowed (since the $ij$ pair start out in a
spin-symmetric combination and cannot
end up occupying the same orbital),
so ${\rm Tr}_1 H^2$ is reduced at the $O(M)$ level. Thus, in the
large-$M$ limit for $C_0=C_1$, we obtain
\begin{equation}
{{\rm Tr}_0 H^2 \over {\rm Tr}_1 H^2} = 1+ O(1/M)\,.
\label{equalcouplingwidth}
\end{equation}
By itself, however, this width
difference should not be sufficient to determine
the ground state behavior. That is
because for a Gaussian spectral density with $d$ states
the location of the ground
state scales as $E_{gs} \sim \sqrt{{\rm Tr} H^2} \sqrt{\log d}$,
while the level spacing near the ground state goes as 
$\rho^{-1}(E_{gs}) \sim \sqrt{{\rm Tr} H^2} /\sqrt{\log d}$.
So the ratio of the expected
difference (gap) between the lowest $S=0$ and $S=1$
states to the level spacing scales as
\begin{equation}
{\log d \over M} \sim {4 \log M \over M} \to 0\,,
\end{equation}
and thus in a model of random uncorrelated fluctuations under the two Gaussian
spectral envelopes $\rho_0(E)$ and $\rho_1(E)$, we would expect no preference
for either $S=0$ or $S=1$ ground states (i.e. the ratio of $S=0$ to $S=1$
ground states should
approach the $3:2$ ratio valid for the {\it total} density of states).
We note, however,
that in our numerical simulations, the ground state always comes out to have
spin $S=0$, even for the highest values of $M$ ($M=12$)
that we can achieve, where
the width difference is very small. Indeed, as we saw above in
Section~\ref{secwidth}, $S=0$ ground states can dominate even when the 
width difference changes sign, as happens for a pure spin-$1$ two-body
coupling ($C_0=0$).
A major reason for this we believe to be
strong deviations from the Gaussian shape of a four-particle spectrum (though
correlations between $\rho_0(E)$ and $\rho_1(E)$ may also be relevant; we
will discuss these correlations in more detail in Section~\ref{correls}).

\subsection{Spectral shape effects}
\label{shape}

We recall that the approach of an $N-$body spectrum to a Gaussian shape given
only two-body interactions depends crucially on the assumption that repeated
applications of the Hamiltonian act on different pairs of particles and therefore
may be taken to commute with each other. So, for example, 
\begin{eqnarray}
\langle {\rm Tr} H^4 \rangle &=& \left\langle \sum H_{\alpha
\alpha'} H_{\alpha'' \alpha'''}
H_{\alpha''' \alpha''} H_{\alpha' \alpha}
\right\rangle \nonumber \\ &+& \left\langle
 \sum H_{\alpha \alpha'} H_{\alpha' \alpha}
H_{\alpha'' \alpha'''} H_{\alpha''' \alpha''}
 \right\rangle \nonumber \\ &+& \left\langle \sum
 H_{\alpha \alpha'}H_{\alpha'' \alpha'''}H_{\alpha' \alpha} H_{\alpha'''
\alpha''}
\right\rangle \nonumber \\ &=& 3\left\langle {\rm Tr} H^2 \right\rangle ^2\,,
\end{eqnarray}
where $\alpha$, $\alpha'$, $\alpha''$, and $\alpha'''$
label two-body states (either $J=0$ or $J=1$), as in Eq.~\ref{twobodyham}, 
and are summed over.
Notice that the third term contributes in general only if the two transitions
involve
distinct pairs of interacting particles. Otherwise (e.g., for a generic
sparse matrix with no two-body interaction structure), we would obtain
$\langle {\rm Tr} H^4 \rangle = 2 \langle {\rm Tr} H^2 \rangle ^2$, the 
relation appropriate to a semicircular spectral shape rather than to a 
Gaussian \cite{Gervots,Mon,Zuker}. So the
extent to which the spectrum deviates from a Gaussian shape towards a 
semicircle may be estimated by considering the probability that successive
two-body interactions do not involve distinct pairs of particles.

Consider for example a pure $J=1$ two-body coupling. In an $S=0$ four-body
state (Eq.~\ref{type0c}), there are a total of three $J=1$ pairs, as we have
already seen above. After two of the particles interact in a two-body $J=1$
combination, the remaining two particles are guaranteed to be also in a
$J=1$ state. So there is a $1/3$ probability that the next interaction
will involve the two particles that have not yet interacted, and thus 
$\langle {\rm Tr}_0 H^4 \rangle \approx (2+{1 \over 3})
 \langle {\rm Tr}_0 H^2 \rangle ^2$. On the other hand, if the total many-body
state is either
$S=1$ (Eq.~\ref{type1c}) or $S=2$, similar counting arguments
lead to the conclusion that there is only a $1/6$ probability that the
second interacting $J=1$ pair will be disjoint from the first; thus
$\langle {\rm Tr}_{1,2} H^4 \rangle \approx (2+{1 \over 6}) 
 \langle {\rm Tr}_{1,2} H^2 \rangle ^2$. This behavior for the spectral
shapes has been numerically confirmed
(e.g., for $M=10$ orbitals,
${\langle {\rm Tr} H^4 \rangle \over \langle {\rm Tr} H^2 \rangle ^2}=
2.42$, $2.18$, and $2.15$, for total many-body spin $S=0$, $1$, and $2$,
respectively), and is similar for
the $J=0$ two-body
interaction. So in particular for the $C_0=C_1$ case of
equal couplings, although the widths of the $S=0$ and $S=1$ spectra approach
each other in the $M \to \infty$ limit (Eq.~\ref{equalcouplingwidth}), 
the fourth moments remain in a finite ratio in this limit
and lead to $\rho_0(E)$ having
a substantially longer tail than $\rho_1(E)$. Therefore,
despite the almost equal widths, ground states in a $C_0=C_1$ system
will be predominantly $S=0$, even in the dilute $M \to \infty$ limit.

\subsection{The general many-body problem}
\label{manybody}

The above analysis of the $N=4$ particle system
generalizes easily to the real $N$-body problem
that we are interested in.
We begin, as before, by addressing the behavior of the spectral widths
in the $S=0$ and $S=1$ spin sectors. Typical basis states in these
two sectors were already given
above in Eq.~\ref{typecmany}.
Now consider a pure $J=0$
coupling. From Eq.~\ref{j0gen} we get
\begin{equation}
{{\rm Tr}_0 H^2 \over {\rm Tr}_1 H^2} = 1 + O\left ({1 \over N^2}\right )
\end{equation}
in the dilute $M \to \infty$ limit. This can be understood as follows.
The only difference scaling as $M^2$
in the spectral width
is that the $ij$ pair in Eq.~\ref{typecmany}
is in a $J=0$ combination for a many body-state of total
spin $S=0$, but not if the many-body state has total spin $S=1$. Thus,
the $S=0$ state has one more $J=0$ pair that can be acted on by the
Hamiltonian, out of a total of $\approx N^2/8$ pairs with $J=0$.
Similarly,
for a pure $J=1$ coupling we have one extra pair that can be acted on
in an $S=1$ many-body state (out of a total of $\approx 3N^2/8$ pairs with 
$J=1$), so finally
\begin{equation}
{{\rm Tr}_0 H^2 \over {\rm Tr}_1 H^2} \approx 1 +{C_0^2-C_1^2 \over
C_0^2+3C_1^2} \times O\left ({1 \over N^2}\right ) \,.
\label{widthmany}
\end{equation}
If we keep the number of particles $N$ fixed and take the system size
$M \to \infty$, then the many-body level spacing of course goes to zero, while
the width ratio remains finite for $C_0 \ne C_1$. Once again we see, this
time for general values of $N$, that assuming the spectral structure is
dominated by the spectral width, we expect $100\%$ ($0\%$) $S=0$ 
ground states if $C_0>C_1$ ($C_1>C_0$), in the dilute $M \to \infty$ limit.

However, we saw in our discussion above of the $N=4$ special case that 
spectral cumulants beyond ${\rm Tr}_S H^2$ need also to be included in a full
analysis. The ratio ${\rm Tr}_S H^4/({\rm Tr}_S H^2)^2$ always favors
a longer tail for $S=0$, so we only need to consider the case of a $J=1$
coupling, where the width effect and the spectral shape effect are competing
with one another.  Referring back to Eq.~\ref{typecmany}, we note
that if particles on orbitals $i$ and $k_z$ interact via a two-body
$J=1$ interaction, then the two particles which remain on sites
$j$ and $l_z$ are also in a $J=1$ state and can subsequently interact 
if and only if the total system is in an $S=0$ state (so that the four
particles $i$, $j$, $k_z$, $l_z$ are in an $S=0$ state). Because
the $j$, $l_z$ pair is distinct from the $i$, $k_z$ pair by construction,
this causes
it to be more likely that the next $J=1$ pair of interacting particles will
be distinct from $i$, $k_z$ if $S=0$, and contributes towards an enhancement
of the shape ratio ${\rm Tr}_0 H^4 / ({\rm Tr}_0 H^2)^2$
as compared with ${\rm Tr}_1 H^4 / ({\rm Tr}_1 H^2)^2$. However, in a many-body
system this effect is suppressed by $O(1/N^3)$ (because there are $O(N^4)$
ways to choose two pairs, and only $O(N)$ of them are involved in this effect).
So for $C_1 > C_0$ the $O(1/N^2)$ difference in widths between 
$\rho_1(E)$ and $\rho_0(E)$ (Eq.~\ref{widthmany})
is expected to exceed the difference in spectral
shapes, leading to a dominance of $S>0$ states near the edge of the spectrum,
in the dilute many-body limit (which we define as
$M \to \infty$ first, followed by $N \to \infty$).

For $C_0>C_1$, on the other hand, both the difference in widths and difference
in spectral shapes favors an $S=0$ ground state, and this behavior is expected
to persist into the dilute many-body limit. For $C_0=C_1$, the relative
width difference vanishes (Eq.~\ref{widthmany}) in this limit, but the
difference in spectral shapes now takes over, again leading to $S=0$
dominance near the edge of the spectrum.

\subsection{Transition at finite densities}
\label{findens}

We have seen that for finite (large enough) number of particles $N$, $S=0$
ground state dominance is expected in the dilute $M \to \infty$ limit, as
long as the spin-zero two-body coupling is at least as large as the spin-one
coupling, $C_0 \ge C_1$. In the contrary case, the situation is reversed,
with higher-spin states dominating behavior near the edge of the spectrum.
What happens, though, if the system is not very dilute? For a finite number
of single-particle orbitals $M$, we must take into account the subleading
behavior in $M$ of the spectral widths. An argument very similar to the
one leading to Eq.~\ref{equalcouplingwidth}, but generalized to large $N$
and to an arbitrary ratio of couplings $C_1/C_0$ shows that the effect
discussed there always favors the $S=0$ width, at  $O(1/NM)$,
independent of which coupling
is stronger. So for $C_0 \ge C_1$ nothing particularly interesting happens
at finite densities; the preference for $S=0$ states near the edge
is only enhanced by finite density effects. This preference, however, becomes
small (compared to the level spacing) in the many-body limit.

On the other hand, for a stronger
spin-$1$ coupling ($C_1>C_0$), we expect the $O(1/N^2)$ effect favoring
$S=0$ ground states and the $O(1/NM)$ effect favoring $S>0$ ground states
to balance each other, leading to a transition in the width ratio
at finite densities. In fact
a closer analysis of the $4-$body system (Eq.~\ref{traces4}), setting equal
the terms proportional to $C_1^2$ in the two expressions, shows that
for a pure $J=1$ coupling we have a transition just below $M=11$, with
a larger $S=0$ width for $M<11$ and larger $S>0$ widths for $M \ge 11$. This
is consistent with our numerical observations above. For comparable coupling
strengths, the transition of course will take place at lower densities; so,
for example for $C_1^2=C_0^2(1+\epsilon)$, the transition in the width ratio
occurs near $M=10/\epsilon$. Although we are unable to observe this
transition numerically for larger numbers of particles, the analytical
arguments suggest that one should take place in general at densities
$N/M \sim \epsilon$ for a small coupling difference
$\epsilon$ and at an $O(1)$ finite density for a pure
$J=1$ interaction.

\subsection{Numerical results}

In Table~\ref{tabgs} we show some results of a numerical calculation for
$N=4$ particles distributed over $5 \le M \le 12$ orbitals, with a pure
$J=1$ two-body coupling ($C_1=1$, $C_0=0$). The first row of the Table
shows the ratio of widths $\sqrt {{\rm Tr}_S H^2}$ in sectors of total
many-body spin $S=0$ and $S=1$. We see that
${\rm Tr}_0 H^2 > {\rm Tr}_1 H^2$ holds
at low $M$ (high density), while in the dilute regime ($M \ge 12$)
the relation is reversed. The crossover occurs near $M=11$ for $N=4$, in good
agreement with our analytical expectations (see Section~\ref{findens}).
The same qualitative behavior 
is predicted and observed for the higher spin sectors; we see from
the second row of the Table that
the crossover between $S=1$ and $S=2$ dominance occurs at $M=9$.

We proceed
to compare the energies of the lowest-lying states in the different spin
sectors, and find a trend towards lower-lying states of higher spin
as $M$ increases, entirely consistent with the results for the spectral widths.
However, by comparing rows 1 and 3 (or 2 and 4) of the Table, we see
that for any given system size $M$, low-lying states with small total
spin are preferred more than would be expected based simply on the width
ratio. Thus, for $M=11$, where the spin-$0$ and spin-$1$ widths are nearly
equal, the ground state of the system is still nearly always $S=0$, and
on average the ground state is observed to have energy $12\%$ lower than
that of the lowest $S=1$ state. This is due largely to the spectral shape
effects discussed above in Section~\ref{shape}, which favor ground states
of lower total spin given equal spectral widths in the different spin sectors.

When comparing $S=0$ and $S=1$ sectors in particular, we recall
from Section~\ref{shape} that the ratio ${\rm Tr}_S H^4 /
\left ({\rm Tr}_S H^2 \right)^2$ is significanlty larger for $S=0$,
for any value of $M$. For $N>4$ (see Section~\ref{manybody}), spectral shape
effects are reduced, but unfortunately we are unable to probe the dilute
$M \gg N$ regime numerically for $N \ge 6$. Instead we stay at $N=4$ particles
and compare the $S=1$ and $S=2$ sectors, which as we saw previously have
nearly equal values of the ratio
${\rm Tr}_S H^4 / \left ({\rm Tr}_S H^2 \right)^2$. Indeed, by
$M=12$ the mean values of the lowest-lying $S=1$ and $S=2$ states become
almost equal, and already by $M=9$ the lowest-lying $S=2$ state appears before
the lowest-lying $S=1$ state in a significant fraction of systems in the
ensemble. The last row of the Table
supports the analytical arguments presented above
for a transition between lower-spin and higher-spin ground states as a function
of the particle density.

\subsection{An alternative model}
\label{secvijkl}

As we have mentioned previously, the model (Eq.~\ref{twobodyham}) assumes
that there is no correlation between two-body interactions in $J=0$
and $J=1$ two-body states. We may instead consider 
an alternative model, motivated by Ref.~\cite{Gervots}
and much subsequent
work, in which the interaction Hamiltonian is separated into a
spin-independent and a spin-dependent part:
\begin{eqnarray}
H&=& \sum_{ijkl}\sum_{J_iJ_jJ_kJ_l} \left(C_V V_{ijkl}+C_W W_{ijkl}
 \vec J_k \cdot \vec J_l \right) \nonumber \\
 &\times& a^\dagger_{iJ_i} a^\dagger_{j J_j} a_{k J_k}
 a_{lJ_l} \delta(\vec J_i + \vec J_j- \vec J_k -\vec J_l) \,,
\label{vijklham}
\end{eqnarray}
where the indices $i,$ $j,$ $k,$ and $l$ label individual orbitals,
$J_i,$ $J_j,$ $J_k,$ and $J_l= \pm {1 \over 2}$ are the $z-$components 
of spin for each of the four fermions, and total spin conservation has been
enforced. The matrix elements $V_{ijkl}$ and $W_{ijkl}$ are taken to
be real Gaussian random and independent variables, with the constraints
\begin{equation}
V_{jilk}=V_{ikjl}=V_{ljki}=V_{ijkl}
\end{equation}
(and similarly for the $W_{ijkl}$) imposed by the reality of wave functions
and the symmetry of the two-body interaction. The constants $C_V$ and
$C_W$ determine, of course, the relative strengths of the spin-independent
and spin-coupled interactions.

The behavior near the ground state for the model of Eq.~\ref{vijklham} can be
understood in terms very similar to those we used for analyzing the original
model of Eq.~\ref{twobodyham}. We assume initially that spin-spin interactions
cannot be neglected, i.e. $C_W \ne 0$.  We first note that the spin-spin
interaction $\vec J_k \cdot \vec J_l = (1/2) ((\vec J_k +\vec J_l)^2 -
J_k^2-J_l^2) = (1/2)((\vec J_k +\vec J_l)^2 -3/2)$ is three times stronger when
particles $k$ and $l$ are in a $J=0$ two-body state as compared with a $J=1$
two-body state. Now we have seen above that an $S=0$ $N-$particle many-body
state has $O(1/N^2)$ more $J=0$ pairs than an $S=1$ many body state: this leads
to a wider $\rho_0(E)$ envelope and $S=0$ ground state dominance, just as for
the $C_0>C_1$ situation considered in the previous model. Thus, a spin-spin
interaction (of either sign) is seen immediately to lead to $S=0$ ground state
dominance ($100\%$ dominance in the $M \to \infty$ many-orbital limit). At
finite densities, $N/M \sim {\rm const}$, we must once again include
corrections to the width difference, of order $1/NM$, as in
Eq.~\ref{equalcouplingwidth} and in Section~\ref{findens} above. We
find, however, that the subleading correction, associated with the
impossibility of doubling up two particles in a $J=1$ two-body state, always
favors the $S=0$ many-body state. This is a model-independent property, 
resulting simply
from the counting of two-body $J=0$ vs. $J=1$ pairs in a many-body state.
Therefore the
$O(1/N^2)$ and $O(1/NM)$ effects contribute with the same sign, and even at
finite density we expect an excess of $S=0$ ground states. Finally, for
small systems, shape effects (discussed above in Section~\ref{shape}) also
contribute, again favoring the $S=0$ ground state.

We may also consider the scenario of a spin-free interaction ($C_W=0$), which
because of the equal strength with which it couples $S=0$ and $S=1$ pairs may
be compared with the $C_0=C_1$ special case discussed above for the model of
Eq.~\ref{twobodyham}. As in that case, the width difference is now expected to
vanish in the $M \to \infty$ dilute limit; furthermore as before the width
difference eventually becomes small compared with the level spacing in the
system. However, at moderate values of the particle number $N$, spectral shape
effects are expected to be very important, leading to $S=0$ ground state excess
at arbitrarily large values of $M$.

\section{Correlations and Pairing}
\label{seccorrelpairing}

In realistic many-body systems there are strong correlations between ground
states of different quantum numbers, and the pairing correlation is of
particular importance. This correlation has also been observed in various
studies of nuclear models with random two-body interactions
~\cite{JBD,BFP,JBDT}. 
In this Section we investigate correlations between
energies and wave functions of low lying states with different quantum numbers.

\subsection{Pairing}
\label{secpairing}

Pairing is a well known phenomenon in many-body fermion systems.  Its spectral
signature is an energy gap that separates the ground state from the excited
states. In the presence of pairing, the ground state is a condensate of pairs
of fermions whose quantum numbers are related to each other by time reversal
symmetry. In the case of the spin systems considered in this work, pairing
refers to a ground state of the form $\prod_j
b^\dagger_{j\downarrow}b^\dagger_{j\uparrow}|0\rangle$, where $b^\dagger_j$
is a quasi-particle operator that is related to the single-particle operators
$a^\dagger_i$ by a unitary transformation, as $b_j=\sum_{\i=1}^M
U_{ji}a_i$.

Let us first consider pairing in the single-particle basis generated by the
$a^\dagger_i$. The results on wave function structure in random spin
systems reported in Ref.~\cite{KP} show that states at high spectral densities
have a number of principal components that agrees with standard random matrix
theory, while low-lying states have a considerably smaller number of principal
components (thus indicating a simpler structure and some degree of regularity
\cite{FyMiIPR,FyMi}). However, the structure even of the low-lying states
was found to be much more complex
than a simple Slater determinant. Based on that previous analysis, and on
the analytical results obtained above in Eq.~\ref{j0gen} 
showing very weak dependence
of spectral width on double occupancy number,
we do not
expect to find pairing here
in the single-particle basis. To compare these arguments with numerical data we
measure the number of doubly occupied single-particle orbitals in a given
eigenstate $|E,S\rangle$ with energy $E$ and total spin $S$ by computing
$P_{E,S}=\langle E,S| \sum_j a^\dagger_{j\downarrow}a^\dagger_{j\uparrow}
a_{j\uparrow}a_{j\downarrow}|E,S\rangle$ (compare the related quantity $D$
for basis states in Section~\ref{secwidthgen}). For an ensemble of
$100$ random Hamiltonians with $N=6$
spin-${1 \over 2}$
fermions on $M=6$ sites, $C_0=1,$ and $C_1=0$, 
$P_{E,S}$ shows fluctuations around $\langle P_{E,0}\rangle\approx
1.71$, $\langle P_{E,1}\rangle\approx 1.43$, $\langle P_{E,2}\rangle\approx
0.86$ and does not exhibit any significant energy dependence. 
Thus, no pairing is present in the single-particle basis $a^\dagger_i$.
The numerical
results agree well with the analytical expressions $P_{E,0}=12/7$,
$P_{E,1}=270/189$, and $P_{E,2}=6/7$.

\begin{figure}
\centerline{
\psfig{file=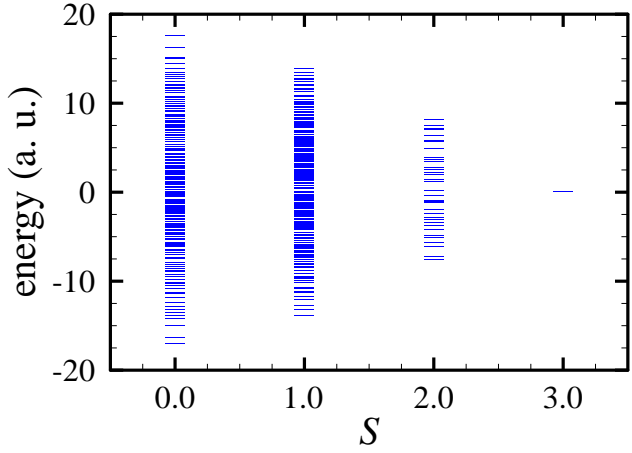,angle=0,width=0.55\textwidth}}
\vskip 0.1in
\caption{Spectra for a system of six fermions on six orbitals with $S_z=0$ 
as a function of total spin $S$.} 

\label{fig_spec}
\end{figure}

Pairing in a general quasi-particle basis can most conveniently be investigated
by examining
its spectral signature. The existence of a pairing gap is closely related to
the approximate degeneracy of the first excited spin-0 level $E_2^{(S=0)}$ with
the lowest-energy spin-one state $E_1^{(S=1)}$. Within the two different models
considered in this work, we do not observe any degeneracy between $E_2^{(S=0)}$
and $E_1^{(S=1)}$, and find the ratio of splitting to gap as
\be
\Delta\equiv\frac{\left|E_2^{(S=0)}-E_1^{(S=1)}\right|}{\min{\left(E_2^{(S=0)},E_1^{(S=1)}\right)}-E_1^{(S=0)}}
 \approx 2.8  
\ee 
for an ensemble of $100$ Hamiltonians with $C_0=1$, $C_1=0$, and
$M=N=6$. Similar results hold for other parameter values, and we never see
$\Delta \ll 1$, which would be indicative of a pairing structure. Thus,
the dominance of spin-0 ground states cannot be caused by pairing correlations
within our model.
Note that this result differs from the
observations made for the nuclear shell model with random
interactions~\cite{JBD,JBDT,BFP}.  
As an illustration we show in Fig.~\ref{fig_spec} the spin dependent spectra 
for a
system of six fermions on six single-particle orbitals. The random two-body
interactions only couple pairs of spin $J=0$, i.e. we set $C_1=0$. The ground 
state and the low energy states have spin $S=0$. The
absence of low-lying spin-degenerate states clearly indicates the absence 
of pairing correlations.

\begin{figure}
\centerline{
\psfig{file=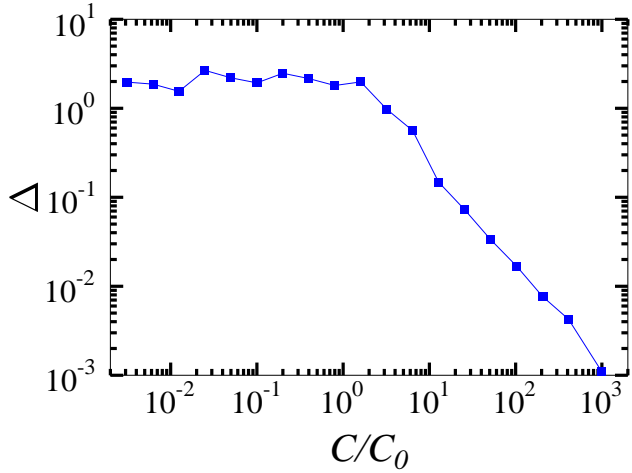,angle=0,width=0.55\textwidth}}
\vskip 0.1in
\caption{Pairing in a many-body system with a random two-body interaction
and a harmonic
one-body potential. $\Delta$ measures the ratio of the splitting 
to the gap, while $C/C_0$ measures the ratio of the one-body 
and the two-body couplings (see text for details).} 
\label{fig_pair}
\end{figure}

To study further the subject of pairing, we add a one-body part of the form 
\be
H_1=C\sum_{j=1}^M
j\left(a^\dagger_{j\uparrow}a_{j\uparrow}+
a^\dagger_{j\downarrow}a_{j\downarrow}\right)
\ee 
to the Hamiltonian (Eq.~\ref{twobodyham}). Fig.~\ref{fig_pair} shows the
pairing observable $\Delta$ as a function of $c\equiv C/(NMC_0)$ for
$C_1=0$ and $N=M=6$.
For $c\gg 1$ we find $\Delta\ll 1$, indicating strong
pairing. In this regime the two-body interaction is a small 
perturbation, and we
obtain pairing in the basis of single particle states. Note that this behavior
is not caused by the harmonic form of our one-body potential but rather is
determined by the applicability of a single-particle picture. For decreasing
values of $c$ one leaves this perturbative regime, and the increase
of $\Delta$ shows that pairing is suppressed. Note that 
$\Delta > 1$ at
sufficiently small values of $c$. This indicates that there are several
spin-0 levels below the spin-1 ground state, as in Fig.~\ref{fig_spec}
above. This finding is consistent with
the results of Section~\ref{secspinstruct} and hints at the dominance of
spin-0 low-lying states. As a check we also
replaced the spin-1 ground state energy by the highest spin-1 energy (which
should not be at all correlated with the spin-0 ground state energy), while
correcting for a nonzero mean of the total spectral density, i.e. replacing
$E_1^{(S=1)}\to {\rm Tr} H-E_{\rm max}^{(S=1)}$,
and recomputed $\Delta$. For the 
system with a pure two-body force we found no difference when compared to the
previous results, confirming that the value of $\Delta$ at $c=0$
is consistent with random spectral fluctuations.

\subsection{Level and energy correlations}
\label{correls}

\begin{figure}
\centerline{
\psfig{file=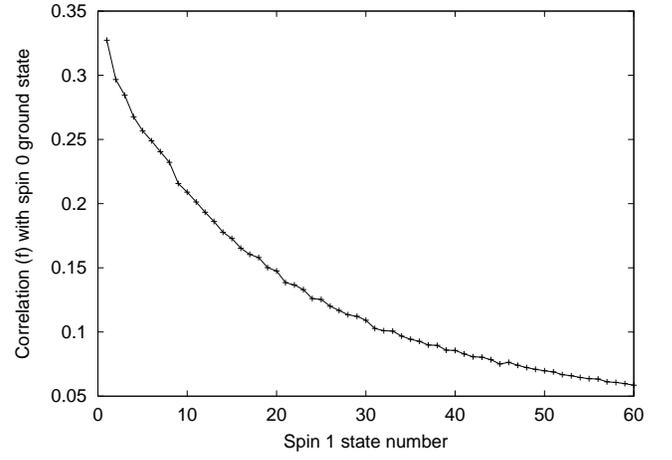,angle=270,width=0.49\textwidth}}
\vskip 0.1in
\caption{Correlation $f$ (see text) between the
spin $0$ ground state and the $n^{\rm th}$ state
of spin $1$. }

\label{fig_corr}
\end{figure}

The absence of pairing in the presence of a strong two-body force is 
peculiar to our model and is
not shared by nuclear shell models with random interactions.
It is therefore interesting to look for different correlations between low
lying states of spin-0 and spin-1. Such correlations have been observed
previously in Refs.~\cite{JBD,JBDT}. We therefore look at the generalized
collectivity defined as
\be
f\equiv {|\langle S=1|B|S=0\rangle|^2\over \langle S=0|B^\dagger B|S=0\rangle},
\ee
where $B=\sum_{i<j} c_{ij} B_{ij}$ and
\be
B_{ij}=
(a_{i \downarrow}^\dagger a_{j \uparrow}^\dagger + 
a_{i \uparrow}^\dagger a_{j \downarrow}^\dagger)
(a_{i \downarrow}a_{j \uparrow} - a_{i \uparrow}a_{j \downarrow})
\ee
is a spin-flip operator that transforms a pair of spin zero into a pair of spin 
one. Following Refs.~\cite{JBD,JBDT} we choose the coefficients as
\be
c_{ij}=\langle S=1|B_{ij}|S=0\rangle.
\ee
This is more convenient than searching for
those coefficients $c_{ij}$ that maximize
$f$. To be definite, we take the ground states of spin zero and spin one,
i.e. $|S=0\rangle=|E_1^{(S=0)}\rangle$ and $|S=1\rangle=|E_1^{(S=1)}\rangle$,
and compare with the corresponding value that is obtained by substituting
$|S=1\rangle \to |E_{\rm max}^{(S=1)}\rangle$.  The states corresponding to the
highest and lowest energies in a sector of definite spin are assumed to be
uncorrelated. The results are presented in Table~\ref{tab1} for $N=4$ and 
different values of $M$, once again setting $C_1=0$.
Obviously, there
is a strong correlation between the ground states of different spin. It is
instructive to let $|S=1\rangle$ run through the $n$ lowest lying spin-$1$
states
$|S=1\rangle=|E_1^{(S=1)}\rangle \ldots |E_n^{(S=1)}\rangle$ while keeping
$|S=0\rangle=|E_1^{(S=0)}\rangle$ fixed. Figure~\ref{fig_corr} shows that the
correlation is strongest for the ground states and decreases for the higher
lying states. This finding is consistent with the recent observation~\cite{KP}
that low-lying states are weakly localized in Fock space, while high-lying 
states are delocalized and agree with random matrix predictions.

It is also interesting to consider energy correlations between ground states 
of spin zero and spin one. The appropriate correlator is
\be
G\equiv
\frac{\left \langle \left (\delta E_0^{(S=0)}
\right)
\left( \delta E_0^{(S=1)} 
 \right) \right \rangle }
{
\left[ \left \langle \left ( \delta E_0^{(S=0)}\right )^2 \right \rangle 
\left \langle \left ( \delta E_0^{(S=1)}\right )^2 \right \rangle 
\right ] ^{1/2}
}
\ee
where $\delta E_0^{(S)} \equiv E_0^{(S)}-\left\langle E_0^{(S)} \right\rangle$
and $\langle \ldots \rangle$ denotes an ensemble average. We find $G=0.93$ 
for $N=4$ particles on $M=9$ sites (with $C_1=0$),
indicating a strong correlation. The observable $G$, is however, somewhat
misleading, since a significant part of the fluctuations $\delta E_0$
from one realization of the ensemble to another arises simply from
the fluctuation of the centroid and width of the overall spectral envelope,
${\rm Tr} H$ and ${\rm Tr} H^2$. The centroid and width for a given system
are almost identical in sectors of different total many-body spin.
To understand fluctuations {\it under} the spectral envelopes, we should
therefore remove this trivial effect by considering the correlation $\tilde G$
of $\tilde E_0^{(S=0)}$ and $\tilde E_0^{(S=1)}$, where
\begin{equation}
\tilde E_0 = {E_0 - {\rm Tr} H \over
\sqrt{{\rm Tr} H^2 - \left({\rm Tr} H\right)^2}   }
\end{equation}
We then obtain $\tilde G=0.76$ for the same set of parameters. The value of
$\tilde G$ increases as one goes to larger systems; e.g. for $N=4$ particles,
$\tilde G$ increases monotonically
from $0.20$ to $0.76$ as $M$ goes from $4$ to $9$.
These observations are consistent
with the existence of correlations between the wave functions of the 
corresponding states.

Note that the observation of strong correlations between low-lying states of
spin-0 and spin-1 is in qualitative agreement with results in shell models and
interacting boson models. However, in contrast with the findings in
these nuclear models,
we do not find any indications of a pairing correlation in the spin systems
considered in this work. One can only speculate that this is due to the 
additional isospin degrees of freedom and the more sophisticated angular 
momentum coupling scheme used in the nuclear models.  

\section{Summary}

We have investigated the spin structure of low lying states in fermion systems
with random two-body interactions. Numerical calculations show that the ground
state has spin zero with almost 100\% probability. For a dilute system we
derive analytical expressions for the shapes and widths of the spectral
densities at different spin. These explain our numerical results.  For systems
with a spin-1 coupling that is stronger than the spin-0 two-body coupling we
predict a transition from spin-0 ground states to ground states of higher
spin. While it is difficult to access this regime numerically, we do observe one
of its precursors -- namely a spin-2 ground state that is lower in energy than
the spin-1 ground state.

Furthermore, we have shown that there are strong correlations between wave
functions and energies of low-lying states of different spin quantum
number. The wave function correlations are manifest in expectation values of
spin transition operators and indicate some degree of collective behavior. The
existence of energy correlations confirm the picture about the dominance of
spin-0 ground states. These correlations decrease as one moves away from the
ground state region. In the absence of a one-body force we find no indication
for any pairing correlation between low lying states, and our model differs
herein from the nuclear models with random interactions. Pairing correlations
are only observed after a sufficiently strong one-body potential is added to
the random Hamiltonian.

\section*{Acknowledgments}
We acknowledge useful discussions with R. Bijker, A. Frank, 
and T. H. Seligman, and we particularly thank G. F. Bertsch for providing
the original motivation and making many helpful suggestions.
This work was partially supported by Department of Energy.

\begin{table}
\begin{tabular}{|c||c|c|c|c|c|c|c|c|c|}
$M$ & 5 & 6 & 7 & 8 & 9 & 10 & 11 & 12 \\ \hline \hline
${\sigma^{(S=0)} / \sigma^{(S=1)}}$ & 1.20 & 1.13 & 1.09 & 1.06 &
 1.03 & 1.01 & 1.00 & 0.99 \\ \hline
${\sigma^{(S=1)} / \sigma^{(S=2)}}$ & 1.21 & 1.12 & 1.07 & 1.03
 & 0.99 & 0.97 & 0.95 & 0.94 \\ \hline \hline
${E_1^{(S=0)} / E_1^{(S=1)}}$ & 1.40 & 1.27 & 1.21 & 1.17 & 1.14
 & 1.11 & 1.12 & 1.09 \\ \hline
${E_1^{(S=1)} / E_1^{(S=2)}}$ & 1.89 & 1.40 & 1.21 & 1.14 & 1.07 &
 1.03 & 1.02 & 1.01 \\ \hline \hline
$E_1^{(S=1)} < E_1^{(S=0)}$ & 0\% & 0\% & 0\% & $0\%$ & 0\% & 0\% &
  0\% & 0\% \\ \hline 
$E_1^{(S=2)} < E_1^{(S=1)}$ & 0\% & 0\% & 0\% & $<1\%$ & 10\% & 20\% &
  20\% & 70\% \\ 
\end{tabular}
\vskip 0.1in
\protect\caption{Relative spectral widths $\sigma^{(S)} \equiv
\sqrt{{\rm Tr}_S H^2}$ and energies of the lowest-lying states
$E_1^{(S)}$ in sectors of total many-body spin $S=0$, $1$, and $2$, for $N=4$
particles on $M$ sites, with a pure $J=1$ two-body coupling. The last row
shows the fraction of systems in the ensemble in which the lowest-lying
$S=2$ state is lower in energy than the lowest-lying $S=1$ state.}
\label{tabgs}
\end{table}

\begin{table}
\begin{tabular}{|c||c|c|c|c|c|c|}
$ M$             & 4 & 5  & 6  & 7  & 8  & 9  \\\hline
$ f$             & 0.45 & 0.37 & 0.33 & 0.29 & 0.24 & 0.26 \\
$f_{\rm random}$ & 0.11 & 0.05 & 0.03 & 0.02 & 0.01 & 0.01 \\
\end{tabular}
\vskip 0.1in
\protect\caption{Wave function correlation $f$ between ground states
of total spin zero and spin one in a system of $N=4$ spin-${1 \over 2}$
fermions on $M$ 
orbitals. $f_{\rm random}$ measures the correlation between the (presumably
uncorrelated) spin-0 ground state and the highest state of spin one (see text
for details).}
\label{tab1}
\end{table}

\end{document}